# THE TITIUS-BODE LAW REVISITED BUT NOT REVIVED

Ivan Kotliarov

197101 do vostrebovaniya

St. Petersburg Russia

E-mail lrpg@mail.ru

## ABSTRACT

The present article gives a detailed analysis of the new formulation of Titius-Bode law by (Poveda, Lara 2008) and of the hypothesis that this law may exist in extra-solar planetary system. A thorough study of the correspondences between the calculated distances and the observed ones in the Solar system and in 55 Cancri is given. It is shown that Poveda-Lara hypothesis contains serious mistakes (both in theory and in calculations) that makes it unacceptable.

## INTRODUCTION

In their article Arcadio Poveda and Patricia Lara (Poveda, Lara 2008) tried to revive the Titius-Bode law (TBL) (Nieto 1972) and to extrapolate it to other planetary systems, namely 55 Cancri. The authors even predicted yet undiscovered planets in that system on a basis of their hypothesis. This approach is indeed very interesting as if this hypothesis have been correct it would be a major step towards proving the physical nature of this highly controversial law. Obviously, if the distribution of planetary distances were governed by TBL not only in the Solar system but also in other planetary systems, it would clearly demonstrate that TBL is something more than a simple numerical coincidence. It would then have to be considered as a strict phenomenological rule waiting for a theoretical basis – just like Kepler's laws before Sir Isaac Newton explained them.

However, this hypothesis has some major flaws that seriously diminish its scientific value. In the present article I will try to discuss these problems.

1. THE STRUCTURE OF THE POVEDA-LARA HYPOTHESIS

The Poveda-Lara hypothesis (PLH) is based on the following assumptions (Poveda, Lara 2008):
1. The traditional form of TBL (1) has to be discarded.

$$a_i = 0.4 + 0.3 \times 2^n, \quad (1)$$

   $a_i$ – semi-major axis of the $i$-th planet (counting from the Sun), AU;

   $n$ – exponent, $n = -\infty$ for Mercury and $n = i - 2$ for all other planets;

2. Instead of the power formula (1) for TBL an exponential one (2) is used:

$$a_n = 0.1912 e^{0.5594n}, \quad (2)$$

   $n$ – orbital number of the planet (counting form the Sun);

3. The formula (2) does not include Mercury and Pluto. Mercury is excluded because its exponent $n$ (the authors call it "the orbital number") in the formula (1) is equal to $-\infty$, which is devoid of any physical meaning and because the value of the constant 0.4 in (1) was arbitrarily chosen in order to ensure correct values of $a_i$ for Mercury and for the Earth. The reasons of Pluto's exclusion are its "pathological" (as the authors say) orbit (the "pathology" is probably the high eccentricity of its orbit – this is why Pluto spends a part of its orbital period within the orbit of Neptune) and the fact that its origin is unknown (it is either a Kuiper belt object or a former satellite of Neptune);

4. The authors suppose that some version of the exponential from of TBL (2) is also valid for 55 Cancri planetary system. They propose the following formula (4) that allegedly gives a good fit for all observed planets in that system:

$$a_n = 0.0142 e^{0.9975n}, \qquad (3)$$

$n$ - orbital number of the planet (counting form the central star);

5. According to the observations, the 55 Cancri system consists of 5 planets. However, the formula (3) gives a good fit with the observed data only if the planet that is now considered to be the fifth is actually the sixth. Therefore the authors predict the existence of a planet between the observed fourth and fifth planets in 55 Cancri. They also predict a seventh planet that should be located beyond the orbit of the fifth observed planet.

## 2. THE "ORBITAL NUMBER" PROBLEM

As one can easily see, Poveda and Lara confuse the meaning of the exponent $n$ in the formulae (1) and (2) – almost certainly due to the fact that two different characteristics are denoted by the same letter $n$. Indeed, in the formulae (2) and (3) proposed by Poveda and Lara $n$ is the orbital number – i.e. the number of the planet counting from the central star. However, it is extremely fallacious to believe that $n$ in the formula (1) is also the orbital number – it is just an exponent there (which *may* be a function of the orbital number). If it had been an orbital number then Poveda and Lara would had been allowed to exclude from the formula (2) not only Mercury, but also Venus, as the orbital number 0 is devoid of any physical meaning too[1] (this fact alone would have sufficed for the authors to realize that $n$ in the formula (1) is not the orbital number at all).

Actually, it would be more correct to write down the traditional formula for TBL (1) as follows:

$$a_n = 0.4 + 0.3 \times 2^N, \qquad (4)$$

$$N = F(n), \qquad (5)$$

$n$ – orbital number.

That is, $N$ is a function of the orbital number, but not the orbital number itself.

---

[1] As we will see below, Poveda and Lara actually did exclude Venus from the formula (2) – but they did not mention it.

As $N = -\infty$ for Mercury and $N = n - 2$ for all other planets I would propose the following explicit presentation of $N = F(n)$:

$$N = \frac{n-2}{\text{sign}(n-1)}. \tag{6}$$

In this case the orbital number for Mercury is 1 – as it is expected to be, but $N = -\infty$ as it is prescribed by TBL in its traditional form.

It is interesting to mention that the formula (6) – while being absolutely obvious – has not been proposed by anybody. Of course, it does not provide the exponent $N$ with a physical meaning as the orbital number is not a physical characteristic of a planet according to the mainstream astronomy. However, it links the unexplained exponent $N$ in the formula (4) with the orbital number of the planet which makes the TBL in its traditional form less esoterical.

3. THE PROBLEM OF THE "EXCLUSION OF PLANETS"

The authors say that they excluded Mercury and Pluto from the formula (2). In my opinion, it would be more correct to say that the authors recognize that Mercury and Pluto do not fit the formula (2) – if these planets had been indeed excluded from that formula then Venus should have been assigned the orbital number 1 (with Mercury excluded from the planetary sequence) which is not the case – the orbital number of Venus is actually 2 in Poveda and Lara's calculations based on the formula (2).

Of course, this is just an incorrect formulation, but the second problem is much more serious – why a formula that should describe the distribution of planetary distances is not valid for all planets? The authors tried to provide us with an explanation of this "exclusion" of Mercury and Pluto. Let us analyze these reasons.

1.  Mercury:

Orbital number of Mercury in the traditional version of TBL is $-\infty$, which has no physical meaning. This problem was discussed in detail in the part 1 of the present paper, and I hope that it is clear for the reader that Poveda and Lara's understanding of the meaning of the exponent in the formula (1) is simply incorrect.

> It is also important to highlight that the explanation of the fact that a parameter (semi-major axis of a planet) does not fit a given statistical formula of semi-major axes of planets is that this parameter corresponds to a strange exponent in another phenomenological formula can hardly be accepted as a valid explanation;

The constant 0.4 in the formula (1) was especially chosen to ensure the better correspondence of calculated distances of Mercury and the Earth to observed ones. The authors probably believe that if one chooses a constant in a phenomenological formula so that calculated parameters correspond to observed ones, it means that the creator of this formula adapts facts to calculations (which is considered to be a serious crime in the scholarly community). Obviously, it is not true: in all empirical formulae constants are chosen in a way that Poveda and Lara would call "arbitral". An empirical formula must provide a good fit between observed and calculated data, and its constants are chosen to ensure this fit. So this objection against including Mercury in the formula (2) must be rejected;

2.   Pluto:

Pluto has a "pathological" orbit – but the authors did not explain why the Pluto's orbit is considered to be "pathological". Probably the "pathology" is the fact that Pluto can come closer to the Sun than Neptune, or, in other words, the high eccentricity of the orbit of Pluto. But if the high eccentricity is an explanation for the absence of correspondence between calculated and real distances for a given planet, then we have to answer two questions: 1) why the eccentricity has such a big importance for a law of planetary distances? 2) if the eccentricity is indeed a reason for "exclusion" of a planet from the formula (2), why this reason was not cited for Mercury with the eccentricity equal to 0.21 – not so far from Pluto's eccentricity (0.25) – this explanation for Mercury's "exclusion" would be much more serious than the reasons cited above? As the authors did not answered these questions, we may safely discard this explanation;

The origin of Pluto is unknown – but the same is true for almost all planets. We know that most planets migrated within the Solar system and the orbital parameters (including distances) underwent changes. Does it mean that a law of planetary distances is allowed to be incorrect for other planets too? Obviously, it does not. Therefore, this explanation should also be rejected.

As one can see, the authors failed to provide us with a plausible explanation of the fact that the formula does not fit Mercury and Pluto. So their "exclusion" from the formula (2) is controversial and looks like to be an *ad hoc* adaptation of facts to a hypothesis.

Interestingly enough, the authors did not bothered to mention if the formula (2) is valid for Eris which is considered to be a dwarf planet and should therefore be included in their formula. We may suppose that Eris is also "excluded" from this formula.

So let us repeat the question: why do we have right to exclude any planet from the formula that is expected to describe the distribution of planetary distances? It seems to me that there is a logical contradiction – either the formula is valid for all planet (including Mercury, Pluto and Eris) or it does not describe the structure of the Solar system. I am afraid that in this case the latter hypothesis is true.

3. THE PROBLEM OF CALCULATED DISTANCES

The authors decided not to provide us with a comparison of distances calculated on a basis of the formula (2) and (3) and distances really observed in the Solar system and in 55 Cancri – they simply published a graph showing that there is a nice correspondence between calculations and observations for the Solar system and indicated that the coefficient of correlation is 0.992 for the formula (2) and 0.997 for the formula (3) (Poveda, Lara 2008) which should be indicative of a non-chance character of these formulae.

However, if we decide to calculate the distances on a basis of the formulae (2) and (3) we will clearly see that the fit is far from being ideal (tables 1 and 2 below).

# TABLE 1
# COMPARISON OF REAL AND CALCULATED DISTANCES
# IN THE SOLAR SYTEM (FORMULA (2))[2]

| $n$ | Planet | $a_R$ (real), AU | $a_P$ (calculated on a basis of the formula (2)), AU | $\Delta = \dfrac{a_P - a_R}{a_R} \times 100\%$ |
|---|---|---|---|---|
| *1* | *Mercury* | *0.387* | *0.335* | *- 13.56* |
| 2 | Venus | 0.723 | 0.585 | - 19.05 |
| 3 | Earth | 1.000 | 1.024 | 2.40 |
| 4 | Mars | 1.524 | 1.792 | 17.57 |
| 5 | Ceres | 2.766 | 3.135 | 13.33 |
| 6 | Jupiter | 5.203 | 5.485 | 5.41 |
| 7 | Saturn | 9.537 | 9.596 | 0.62 |
| 8 | Uranus | 19.191 | 16.790 | - 12.51 |
| 9 | Neptune | 28.263 | 29.376 | 3.94 |
| *10* | *Pluto* | *39.482* | *51.396* | *30.18* |
| <u>11</u> | <u>Eris</u> | <u>67.668</u> | <u>89.924</u> | <u>32.89</u> |

The discrepancies are very high for Venus, Mars, Ceres[3] and Uranus. Unfortunately, authors did not provide us with any good explanation of these differences between the formula and the reality. They do not even mention them.

It is important to indicate that $\Delta$ is huge for Venus and Mars, which, according to the authors, fit the formula. However, Venus and Mars are actually "excluded" from the formula (2) – even Mercury fits this formula better than Venus and Mars

---

[2] Data for planets which are, according to Poveda and Lara, "excluded" from the formula (2) are given in italics. Data for Eris, which is not even mentioned by Poveda and Lara, are underlined.

[3] The situation with Ceres may be fixed if one supposes that Ceres has not be necessarily chosen as the typical representative of the asteroid belt: it would suffice to replace it with another massive body from the main belt (the best fit would be ensured with 52 Europe, $a_R = 3.101$, which is the sixth largest asteroid). But in this case one has to explain why the formula of distribution of *planetary* distances is not valid for a dwarf *planet* which is replaced by another massive body from the main belt?

do. It is surprising that the authors did not indicate this fact and did not try to explain it.

It is not necessary to highlight that if we accept such big discrepancies then we are allowed to use virtually *any* formula to describe distribution of planetary distances. Unfortunately, the same discrepancies between calculated and observed distances exist in 55 Cancri, too:

TABLE 2

COMPARISON OF REAL AND CALCULATED DISTANCES IN CANCRI 55[4]

| $n$ | $A_R$ (real), AU | $A_P$ (calculated on a basis of the formula (3)), AU | $D = \dfrac{A_P - A_R}{a_R} \times 100\%$ |
|---|---|---|---|
| 1 | 0.038 | 0.039 | 2.63 |
| 2 | 0.115 | 0.104 | - 9.57 |
| 3 | 0.24 | 0.283 | 17.92 |
| 4 | 0.781 | 0.768 | - 1.66 |
| 5 | ? | *2.08* | ? |
| 6 | 5.77 | 5.643 | - 2.20 |
| 7 | ? | *15.3* | ? |

As we can see, $D$ is high for $n = 2$ and $n = 3$, which may be indicative of a problem with the formula for TBL (3). Again, $D$ for the 3$^{rd}$ planet is higher than for Mercury, so the authors should have "excluded" this planet from the formula (3).

The authors did not provide any explanation for these differences either. Neither did they even indicate them.

Interestingly enough, despite the high discrepancies between theoretical and observed data for known planets, the authors predict the precise distances for yet unknown planets – not the range of distances where these planets should be looked for.

---

[4] Calculations for predicted planets are given in italics.

One can easily see that the difference between the observed data and the distances calculated on a basis of the formulae (2) and (3) is very high, which allows us to reject them.

4. METHODOLOGICAL PROBLEMS

It seems to me that the authors overlooked two very important problems in their methodology of prediction of new planets in other planetary systems (even if we admit that there is a fit between real and theoretical data for "included" planets):

1. As the authors admit themselves, the formula (2) is "truncated" – it "excludes" two planets. In my opinion, it is incorrect to use methodology that requires "excluding" planets for precise prediction of new planets in a different planetary system: this methodology is just unable to ensure the necessary degree of precision – we must always keep in mind that unknown planets may just be anywhere, at any distance form the central star (at least within the orbit of the first known planet and beyond the orbit of the last known one);

2. It is well known that an exponential formula of planetary distribution in the Solar system either "excludes" some planets or creates gaps between known planets (Badolati 1982). The formula (2) is a nice example of "exclusion". And we are allowed to suppose that the planets "predicted" by the formula (3) are just an example of such gaps. At least the authors should have bothered to show the difference between their predictions and the gaps between Vesta and Mars and between Saturn and Uranus predicted by the Armellini-Basano-Hughes law $a_n = 0.283 \times 1.52^n$ (Badolati 1982).

So from the methodological point of view it is incorrect to use the authors' model for precise prediction of new planets and the authors could not effectively predict new planets in 55 Cancri – they just introduced gaps between known planets in that system. These gaps may be occupied but yet undiscovered planets – but not necessarily. And the authors did not try to show why the hypothesis of existence of new planets in these gaps should have the priority.

CONCLUSION

TBL is a very important issue in the planetary science as it is close to the status of a phenomenological law but has no theoretical explanation (or at least it has no explanation accepted by the majority of the scholarly community). An attempt made by Poveda and Lara to confirm this law by the data from a different planetary system is very interesting and important as it could have helped us to better understand the nature of TBL. However, due to serious mistakes committed by the authors their hypothesis should be rejected and the question of existence of TBL in other planetary systems (as well as the question of its best mathematical form in the Solar system) remains open.

REFERENCES

1. Badolati, E. 1982. A supposed new law for planetary distances. In *The Moon and the planets*, **26**.
2. Nieto M. 1972. *The Titius-Bode law of planetary distances: its history and theory*. Oxford.
3. Poveda, A., Lara, P. 2008. The exo-planetary system of 55 Cancri and the Titius-Bode law. *Revista Mexicana de Astronomía y Astrofísica*, **44**, 1, http://www.astroscu.unam.mx/~rmaa/rmaa.html, astro-ph/0803.2240v1.